\documentclass[10pt, a4paper]{article} \usepackage{lrec}
\usepackage{multibib} 
\newcites{languageresource}{Language Resources}
\usepackage{graphicx} 
\usepackage{tabularx} 
\usepackage{booktabs} 
\usepackage{soul}
\usepackage[utf8]{inputenc}
\usepackage{hyperref} 
\usepackage{xstring}

\graphicspath{{./}{figures/}}

\title{A Method for Analysis of Patient Speech in Dialogue for Dementia Detection} %
\name{Saturnino Luz, Sofia de la Fuente, Pierre Albert}
\address{Usher Institute of Population Health Sciences \&
  Informatics\\
  Edinburgh Medical School\\
  The University of Edinburgh, Scotland, UK\\ 
  \{s.luz,sofia.delafuente,pierre.albert\}@ed.ac.uk\\}

\abstract{%
  We present an approach to automatic detection of Alzheimer's type
  dementia based on characteristics of spontaneous spoken language
  dialogue consisting of interviews recorded in natural settings. The
  proposed method employs additive logistic regression (a machine
  learning boosting method) on content-free features extracted from
  dialogical interaction to build a predictive model.  The model
  training data consisted of 21 dialogues between patients with
  Alzheimer's and interviewers, and 17 dialogues between patients with
  other health conditions and interviewers. Features analysed included
  speech rate, turn-taking patterns and other speech
  parameters. Despite relying solely on content-free features, our
  method obtains overall accuracy of 86.5\%, a result comparable to
  those of state-of-the-art methods that employ more complex lexical,
  syntactic and semantic features. While further investigation is
  needed, the fact that we were able to obtain promising results using
  only features that can be easily extracted from spontaneous
  dialogues suggests the possibility of designing non-invasive and
  low-cost mental health monitoring tools for use at scale.
  \\ \newline%
  }

\begin{document}

\maketitleabstract

\section{Introduction}
\label{sec:introduction}

Research into early detection of Alzheimer's disease (AD) has
intensified in the last few years, driven by the realisation that in
order to implement effective measures for secondary prevention of
Alzheimer's type dementia (ATD) it may be necessary to detect AD
pathology decades before a clinical diagnosis of dementia is made
\cite{bib:RitchieCarriereEtAl17al}.  While imaging (PET, MRI scans)
and cerebrospinal fluid analysis provides accurate diagnostic methods,
there is an acknowledged need for alternative, less invasive and more
cost-effective tools for AD screening and diagnostics
\cite{bib:LaskeSohrabiEtAl15inal}. A number of neuropsychological
tests have been developed which can identify signs of AD with varying
levels of accuracy
\cite{Mortamais2016,bib:RitchieCarriereEtAl17al}. However, the
proliferation of technologies that enable personal health monitoring
in daily life points towards the possibility of developing tools to
predict AD based on processing of behavioural signals.

Speech is relatively easy to elicit and has proven to be a valuable
source of clinical information.  It is closely related to cognitive
status, having been used as the primary input in a number of
applications to mental health assessment. It is also ubiquitous and
can be seamlessly acquired. In recent years, combinations of signal
processing, machine learning, and natural language processing
have been proposed for the diagnosis of AD based on the
patient's speech and language
\cite{bib:FraserMeltzerRudzicz16lfidal}. Models built on phonetic,
lexical and syntactic features have borne out the observation that
these linguistic processes are increasingly affected as the disease
progresses \cite{bib:Kirshner12ppapal}. However, most machine learning
research in this area has employed either recorded narrative speech
\cite{Lopez-De-Ipina2012}, or recorded scene descriptions
\cite{bib:LuzCBMS17,bib:FraserMeltzerRudzicz16lfidal} collected as
part of a neuropsychological assessment test, such as the
Boston ``cookie theft'' picture description task
\cite{bib:BeckerEtAllPittCorpus94}.

In contrast to those methods, our approach employs spontaneous
conversational data, exploring patterns of dialogue as basic input
features. Content-free interaction patterns of this kind were first
used in the characterisation of psychopathology by Jaffe and Feldstein
\shortcite{bib:JaffeFeldstein70}, who represented therapist-patient
dialogues as Markov chains. Here, we build on these ideas to analyse
patient data from the Carolina Conversations Collections (CCC)
\cite{pope2011finding}. We trained machine learning models on these
data to differentiate AD and non-AD speech. This work is, to the best
of our knowledge, the first to employ low-level dialogue interaction
data (as opposed to lexical features, or data from narrations other
forms of monologue) as a basis for AD detection on spontaneous speech.

\section{Background}
\label{sec:background}

One of the greatest challenges facing developed countries, and
increasingly the developing world, is the challenge of improving the
quality of life of older people. In 2015, the First Ministerial Conference of
the WHO on Global Action Against Dementia estimated that there are
47.5 million cases of this condition in the world. Cohort studies show
between 10 and 15 new cases per each thousand people every year for
dementia, and between 5 and 8 for Alzheimer's Disease. Prognosis is
usually poor, with an average life expectancy of 7 years from diagnosis.
Less than 3\% diagnosed live longer than 14 years. Current statistics
predict that the population aged over 65 is expected to triple between
years 2000 and 2050 \cite{world2015first}. This will lead to
structural and societal changes, accentuating what is already becoming a highly
demanding issue for health care systems.

Dementia is therefore set to become a very common cause of disability
which places a heavy burden on carers and patients alike. While there
are currently neither a cure nor a way to entirely prevent the
progress of the disease, it is hoped that a better understanding of
language and communication patterns will contribute to
secondary prevention. A characterisation of communication patterns
and their relation to cognitive functioning and decline could be
useful in the design of assistive technologies such as adaptive
interfaces and social robotics \cite{Therapy2008}. These technologies
might help provide respite to carers, and stimulate cognitive,
physical and social activity, which can slow disease progression and
improve the patient's quality of life \cite{Middleton2009}. Collecting
relevant real life observational data and assembly of prior and
current knowledge \cite{Therapy2008} could lead to new effective and
personalised interventions.

Assessing people's behaviour in natural settings might also contribute
to earlier detection \cite{Parsey2013,Mortamais2016}. Language
impairment is a common feature of dementia, implying signs such as
word-finding and understanding difficulties, blurred speech or
disrupted coherence
\cite{AmericanPsychiatricAssociation2000}. Although language is a good
source of clinical information regarding cognitive status, manual
analysis of language by mental health professionals for diagnostic purposes is
challenging and time-consuming. Advances in speech and language
technology could help by providing tools for detecting reliable
differences between patients with dementia and controls
\cite{Bucks2000}, distinguishing among dementia stages
\cite{Thomas2005} and differentiating various types of dementia
\cite{bib:FraserMeltzerRudzicz16lfidal}.

Features such as grammatical constituents, vocabulary richness,
syntactic complexity, psycholinguistics, information content,
repetitiveness, acoustics, speech coherence and prosody, have been
explored in conjunction with machine learning methods to identify
Alzheimer's and other types of dementia through the patient's
speech. This is not only because language is impaired in these
patients, but also because language relies on other cognitive
functions, such as executive functions, which allow us to interact in
a sound and meaningful way. These functions are responsible for
decision making, strategy planning, foreseeing consequences and
problem solving, which are essential to successful communication, but
are impaired by AD
\cite{bib:FraserMeltzerRudzicz16lfidal,Marklund2009,Satt2013}. Although
hardly perceptible to the speakers themselves, patterns of impairment
are thought to occur even in informal and spontaneous conversations
\cite{Bucks2000,Lin2012}.

Our hypothesis in this paper is that people with an AD diagnosis will
show identifiable patterns during dialogue interactions. These
patterns include disrupted turn taking and differences in speech
rate. These indices relate to the fact that, in general, patients with AD 
show poorer conversation abilities and their normal
turn-taking is repeatedly interrupted. Therefore, we expect less
conversational fluidity overall in the AD group dialogues, as compared
to non-AD group. Our approach, which does not rely on transcription
but only on speech-silence patterns and basic prosodic information,
obtains levels of accuracy comparable to state-of-the-art systems that
rely on more complex feature sets.

\section{Related work}
\label{sec:related-work}

Potential applications of the kind of speech technology described in
this paper include the development of interactive
assistive technologies, and monitoring of users for signs of
cognitive decline with a view to mitigating further decline.

From the perspective of potential applications of automatic speech
analysis to technology-assisted care, there is evidence
\cite{Rudzicz2014} that it is psychologically more acceptable for a
user to be aided by another person or a robot than from ambient
sensors and devices which are unable to offer meaningful
interaction. Therefore, the development of such assistive applications
involves research on speech processing for natural conversations
rather than scripted speech or monologues \cite{Conway2016}.

From the perspective of monitoring for early detection, it is known
that AD leads to disruption of one's ability to follow dialogues, even
in simple, routine interactions. At later stages of the disease,
failure to perform meaningful interactions appears
\cite{watson1999analysis}. This has a negative impact on tasks such as
following instructions regarding household activities and medication,
as well as preventing rewarding social interactions. Here, once again
the focus should be on natural interaction data, as scripted talk
cannot be compared to spontaneous conversation in terms of information
richness and external validity of results \cite{Kato2013}. Over the
last decades, different approaches have targeted early detection of
AD on spontaneously generated data through automatic and non-invasive
intelligent methods. Some of these approaches have focused on speech
parameters analysis: automatic spontaneous speech analysis (ASSA),
emotional temperature (ET), \cite{Lopez-De-Ipina2012}, voiceless
segments, and phonological fluency have been shown to explain
significant variance in neuropsychological test results
\cite{GarciaMeilan2012}. These methods are not only non-invasive and
free from side-effects, but also relatively cheap in time and in terms
of resources. Another approach that rely on easily extracted acoustic
features, such as the ones we propose in this paper, though not in
dialogical or spontaneous speech settings is presented by Satt et
Al. \shortcite{Satt2013}. This approach extracts a number of voice
features (voiced segments, average utterance duration, etc.) from
recordings of picture description, sentence repetition, and repeated
pronunciation of three syllables used in diadochokinetic tests in
succession. The method achieves accuracy levels of over 80\% in
detection of AD and mild cognitive impairment (MCI).

Other approaches have used time-aligned transcripts and syntactic
parsing, extracting speech features and using them for classifying
healthy elderly subjects from subjects suffering AD or MCI, as well as
other tasks. This classification has been done either by comparing
impaired to healthy speech performance (speech quality in terms of
lexicon, coherence, etc.), or by comparing classifier performance when
only neuropsychological tests are included against performance when
such tests are used together with speech features, generally with
statistically significant improvements
\cite{Roark2011,bib:FraserMeltzerRudzicz16lfidal}.

Analysis performed on similar corpora provide good insight of the
performances achieved using different features. A first analysis
\cite{bib:FraserMeltzerRudzicz16lfidal}, based on a monologue corpus (DementiaBank),
identified four different linguistic factors as main descriptors:
syntactic, semantic, and information impairments, and acoustic
abnormality. They achieved accuracy of up to 92.05\% using full scale
analysis of 25 features, selected amongst an original feature set of
370 features after extensive experimentation.

An analysis of the CCC corpus by Guinn et al \cite{guinn2012language}
used similar linguistic features. Unlike the work presented in this
paper, Guinn's analysis was focused on the differences between
interviewers and subjects in the subset of patients with AD. They
achieved a combined accuracy of 75-79.5~\% using decision trees, with
a large discrepancy between AD (38-42~\%) and non-AD (74-100~\%)
recognition accuracy.

Works on dialogue so far have identified features such as
conversational confusion (AD increases confusion rates, and this
relates to slower and shorter speech; \cite{Rudzicz2014a}, prosodic
measures \cite{Gonzalez-Moreira2015}, and emotion
\cite{Devillers2005}. These studies used machine learning methods
(neural networks, Na\"{i}ve Bayes, and random forests, respectively),
reporting accuracy in the 70-90~\% range. Although these results are
promising, they are difficult to generalise. This is because they are
primarily content dependent. That is, they employ lexical, and
sometimes syntactic information, which present a number of potential
disadvantages. The content of a conversation is likely to change
greatly depending on whether a participant belongs to the control
group or to the group with Alzheimer's Disease, especially if the
conversational partner is their doctor. In addition, such content is
difficult to acquire in spontaneous speech settings. Despite the
advances in automatic speech recognition, recognition (word) error
rates in unconstrained settings are still over 11\%, even for fairly
clear, telephone dialogues \cite{bib:XiongDroppoEtAl16achpc}. Another
difficulty with these approaches is the fact that they are
language-dependent, and therefore require building different models
for different languages, which in the context of global mental health
could be a major shortcoming. Therefore, these models should aim to
be as content-independent as possible to be generalisable
\cite{Satt2013}. In contrast to content-based approaches, our method
focuses on the interaction patterns themselves, rather than on
characteristics of the speech and language content as such.

\section{Methods}

\subsection{Dataset}

We have conducted our analysis using the Carolina Conversations
Collection \cite{pope2011finding}. The dataset is a digital collection
of recordings of conversations about health, including both audio and
video data, with corresponding transcriptions. The corpus consists of
natural conversations involving an older person (over the age of 65)
with a medical condition. Several demographic and clinical variables
are also available, including: age range, gender, occupation prior to
retirement, disease diagnosed, and level of education (in years). The
interviewers were gerontology and linguistic students or researchers
to whom the patients spoke at least twice a year. A unique alias was
assigned to each patient to protect their identity.

Access to the data was provided after complying with the ethical requirements
of the University of Edinburgh and the Medical University of
South Carolina. In order to ensure that the results described here are reproducible we will
provide, on request, the identifiers for the dialogues used in our
experiments so that interested researchers can recreate our dataset
upon being granted access to the CCC. The source code used for
processing the data is available at a University of
Edinburgh gitlab server\footnote{https://cybermat.tardis.ed.ac.uk/pial/CCCdataset}.

For the research described here we selected a total of 38 patient
dialogues: 21 patients had a diagnosis of Alzheimer's disease (15
females, 6 males), and 17 patients (12 females, 5 males) had other
diseases (diabetes, cardiac issues, etc., excluding neuropsychological
conditions), but not AD. These groups were selected for matching age
ranges and gender frequencies so as to avoid statistical bias. The
dataset also included time-aligned transcripts, which we did not use
except for the computation of an alternative speech rate feature as
described below.

\subsection{Data Preparation}
\label{sec:data-preparation}

\begin{figure}[tb] \centering
  \includegraphics[width=\linewidth]{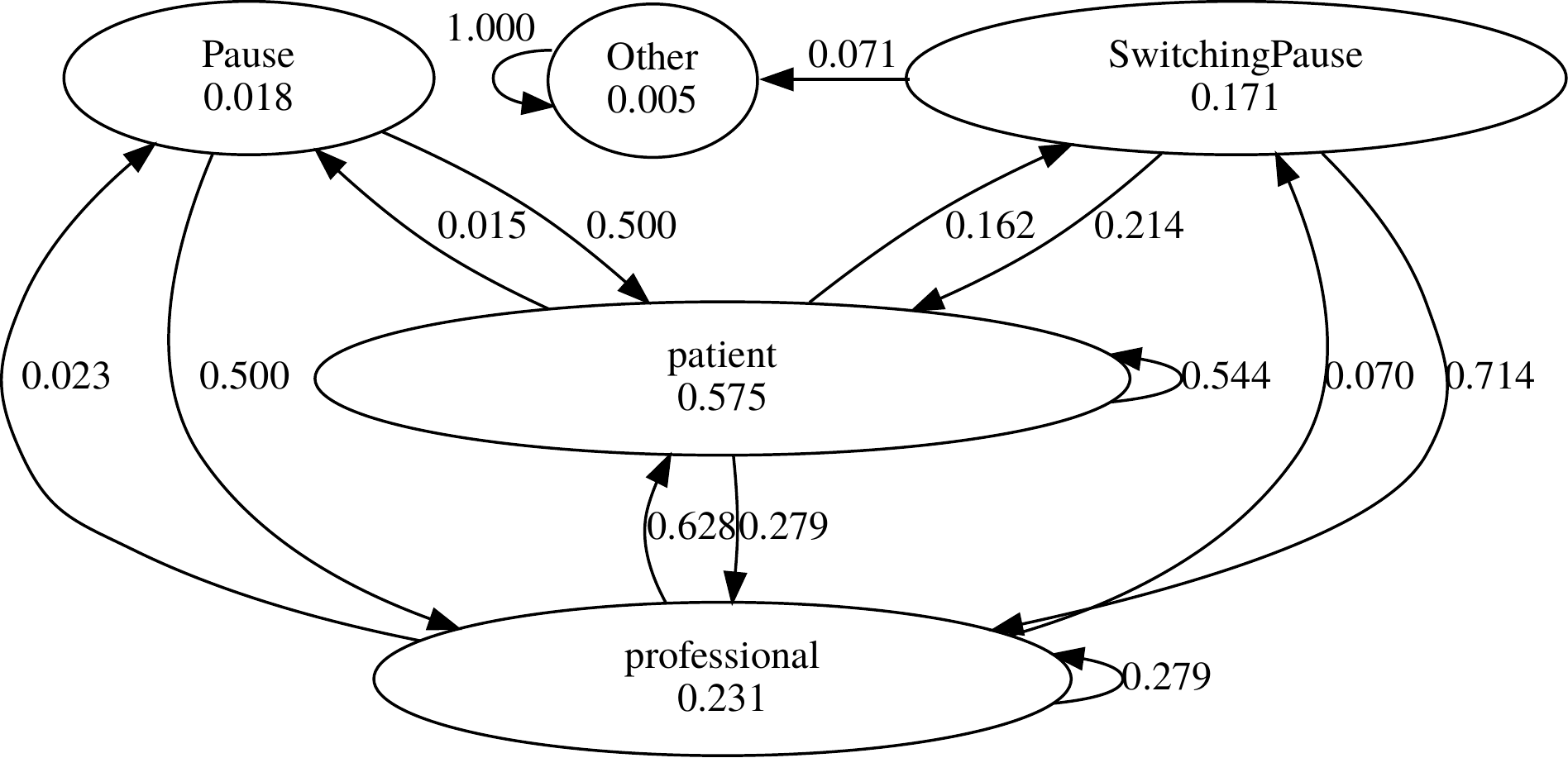}
  \caption{Vocalisation diagram for a patient dialogue.}
  \label{fig:vocgraph}
\end{figure}

The speech data selected as previously described were pre-processed in
order to generate {\em vocalisation graphs} --- that is, Markov
diagrams encoding the first-order conditional transition probabilities
between vocalisation events and steady-state probabilities
\cite{Luz2013}.
Vocalisation events are classified as speech by either the patient or
the interviewer/others, joint talk (overlapping speech), or silence
events (also known as 'floor' events, which are further in the
diagrams as pauses and switching pauses, according to whether the
floor is taken by the same speaker or another speaker, respectively).
An example of vocalisation graph is shown in
Figure~\ref{fig:vocgraph}. 

Vocalisation and pause patterns have been successfully employed in the
analysis of dialogues in a mental-health context
\cite{bib:JaffeFeldstein70}, segmentation \cite{bib:LuzSu10rtpov} and
classification of dialogues, and more recently on characterisation of
participant role and performance in collaborative tasks
\cite{Luz2013}. Models that employ basic turn-taking statistics have
also been proposed for dementia diagnosis
\cite{bib:MirheidariBlackburnEtAl16d}, though not in a systematic
content-fee framework as in our proposed method.

\begin{figure}[htb]
  \centering
  \includegraphics[width=\linewidth]{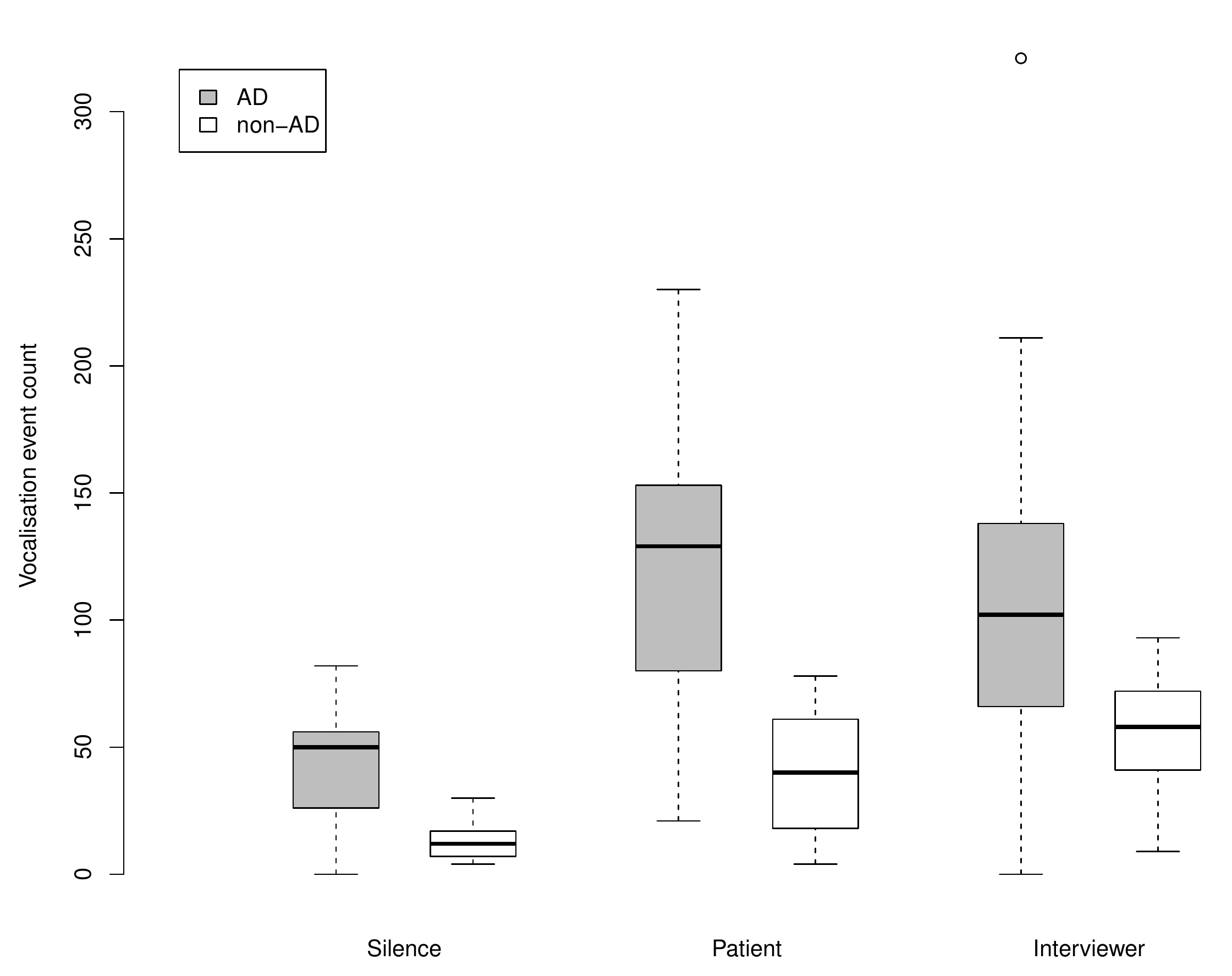}
  \caption{Distribution of vocalisation event counts for patients with and without AD in CCC dialogues.}
  \label{fig:vocboxplot}
\end{figure}

The distributions of event counts according to vocalisation events is
shown in Figure~\ref{fig:vocboxplot}. It can be observed that patients
with AD tend to produce more vocalisation events than their
interviewers (and, consequently, produce more silence events). This is
consistent with findings in the literature on language changes in AD
\cite{AmericanPsychiatricAssociation2000}.

Speech rate was estimated using De Jong's syllable nuclei detection
algorithm \cite{bib:JongWempe09p}, which is an unsupervised method --
that is, it can be applied directly to the acoustic signal, with no
need of human annotation. However, as the audio quality of the CCC
recordings is uneven, and as the dataset provides no gold standard
against which one could assess syllable count, we decided to validate
the use of De Jong's method against the time-stamped transcripts
provided.  Using these transcripts one could, in principle, estimate
average words per minute (WPM) for individual utterances, as is
sometimes done \cite{bib:AkiraLuzIS2017}.  However, this method  of
measuring WPM based on transcription has a number of limitations. Words
have variable length, and their articulation can vary greatly
due to a number of speech-related phenomena, such as phonological
stress, frequency, contextual predictability, and repetition
\cite{bell_predictability_2009}.  In order to mitigate these problems,
we instead produced {\em speech rate ratio} estimates normalised through
a speech synthesizer, employing the methods proposed by Hayakawa et
al. \newcite{bib:AkiraLuzIS2017}. These estimates represent deviations
from a ``normalised'' pace of 160 words per minute (WPM) synthesised
using the MaryTTS system \cite{schroeder2003}. We therefore computed
the ratio of the synthesised speech to the actual duration of the
patient's speech. 
The speech rate ratio correlated well with the syllable per minute
rate extracted using only the recorded audio
($\rho=0.502$,
$t(30)=3.19,
p=0.003$) indicating that speech rate can be estimated with an
acceptable level of reliability through the unsupervised method, even
in fairly noisy settings.

A Python script was employed to extract basic
speaker turn time stamps, speaker role information, and transcriptions
from the original XML-encoded CCC data. The resulting data were then
processed using the R language in order to detect silence intervals,
and categorise turn transitions and pause events. 

Some descriptive statistics on the dialogues can be seen in
Table~\ref{tab:desc_stats}. These statistics include: average turn
duration (how many seconds a participant speaks on average), total
turn duration (how many seconds did the participant's turns lasted in
total), normalised turn duration (the ratio of a participant's turn
duration to the total duration of AD or non-AD dialogues, according
the participant's class), number of words generated (total per class
and on average per class' participant), and number of words per minute
(average per class participant).

\begin{table}
\caption{Descriptive statistics on dialogue turn-taking (duration
given in seconds).}
\label{tab:desc_stats}
\begin{tabular}{lrr}
\toprule
Feature &        non-AD &            AD \\
\midrule
Dialogue duration               &  4107.3 &   7628.4 \\
Dialogue duration TTS            &  7618.8 &   7618.8 \\
Avg turn duration            &    97.3 &    255.8 \\
Total turn duration          &  1654.3 &   4348.3 \\
Norm. total turn duration     &     3.0 &      4.1 \\
Avg turn duration TTS        &   107.6 &    238.0 \\
Total turn duration TTS      &  1829.7 &   4046.1 \\
Norm. total turn duration TTS &     3.0 &      4.2 \\
Avg number of words            &   314.6 &    742.5 \\
Total number of words          &  5348.0 &  12622.0 \\
Avg words per minute                 &   155.9 &    166.5 \\
\bottomrule
\end{tabular}
\end{table}
Contrary to our expectations, we did not observe a statistically
significant difference between the speech rate in syllables per minute
between patients with and without AD (Welch two sample t-test
$t(30.5)=1.15, p = 0.28$), even though the mean for non-AD ($M=180.8$
syllables/min, $sd=28.4$) was higher than that for patients with AD 
($M=168$ syllables/min, $sd=35.6$).

Two alternative data representations were generated. The first
(henceforth referred to as VGO) was based on the vocalisation graphs
only. That is, VGO encodes the probabilities of each possible pair of
transitions, including self-transitions, which tend to dominate Markov
chains sampled, and the steady-state probabilities for each
vocalisation event. The second form of representation (VGS) simply
consists of the VGO with information about the participant's speech
rate (mean and variance) added to the vocalisation statistics. With
the exception of speech rate ratio, which necessitates transcription, all
the information needed to build VGO and VGS instances can be extracted
through straightforward signal processing methods.   

\subsection{Machine learning}

The data instances in the two alternative representation schemes were
annotated for presence or absence of Alzheimer's Disease (AD). A
supervised learning procedure was employed in order to train
classifiers to predict such annotations on unseen data.

We trained a boosting model \cite{bib:SchapireFreund14b} using
decision stumps (i.e. decision trees with a single split node) as weak
learners. The training process consisted of 10 iterations whereby, for
each training instance ($x_i$), a weak classifier $\hat f_m$ was
fitted using weights on the data which were iteratively computed so
that the instances misclassified in the preceding step had their
weights increased by a factor proportional to the weighted training
error. In this case class probability estimates $P(ad=1|data)$ were
used to compute these weights and to weigh the final classification
decision (additive logistic regression) following the Real Adaboost
algorithm \cite{bib:FriedmanHastieTibshirani00ad}:

\begin{eqnarray}
 \label{eq:logit}  \hat F(x) & = & sign\left[ \sum_{m=1}^M \hat f_m(x)\right ]
\end{eqnarray}

Classification performance was assessed through a 10-fold cross
validation procedure.  As the dataset is reasonably balanced, results
were assessed in terms of accuracy, precision (the ratio of the number
of true positives to the number of instances classified as AD), recall
(or sensitivity, the ratio of true positives to the number of AD
cases) and $F_1$ score (the harmonic mean of precision and
recall). Micro ($\mu$) and macro ($M$) averages for these scores are
given by taking means over the entire set of classification decisions
and over individual classifiers respectively, across the 10 folds.
As the data set is fairly small, we also ran a leave-one-out cross
validation (LOOCV) procedure to obtain better estimates of
generalisation accuracy. This consisted of selecting one instance for
testing, and building a classification model on the remaining
instances, and iterating this procedure until all instances have been
selected as testing instances. Macro averages are uninformative in
LOOCV, so we only report overall accuracy figures for this procedure.

ROC curves showing the relationship between true positive and false
positive rates as the classification threshold is varied were also
plotted. Simulation was employed in order to smooth these ROC curves by
running 10 rounds of 10-fold cross validation tests with a randomised
selection of instances making up the hold-out sets.

\section{Results}

\begin{figure}[htb] \centering
  \includegraphics[width=\linewidth]{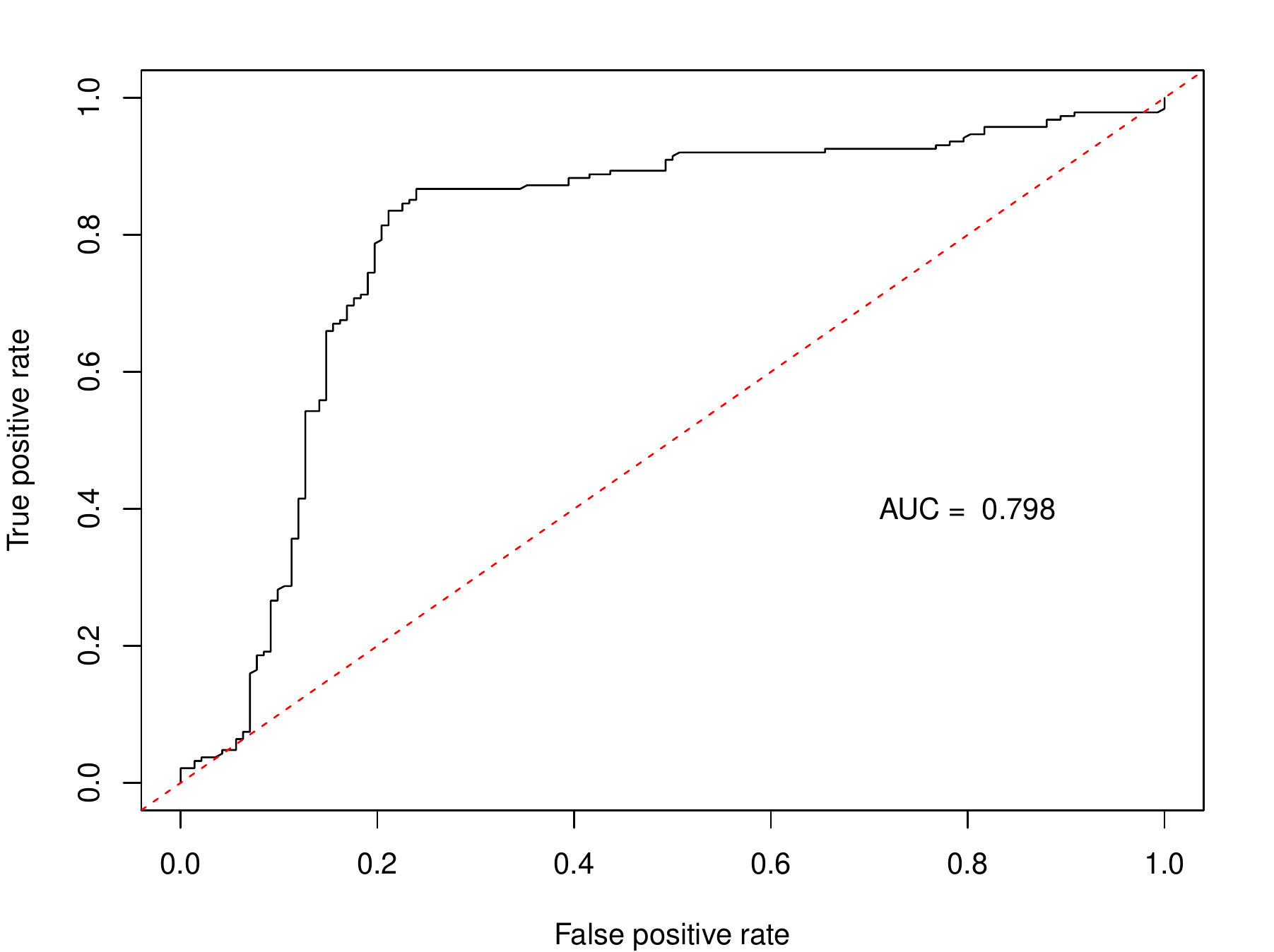}
  \caption{ROC curve for VGO-based classifiers.}
  \label{tab:vgoroc}
\end{figure}

Our first approach, based on the VGO data representation scheme,
produced promising results. Accuracy levels were well above the
baseline, with overall accuracy reaching 81.1\%, showing that
turn taking patterns can provide useful cues to the detection of AD in
dialogues. The results for the VGO-based classification are shown in
Table~\ref{tab:vgoresults}. The corresponding ROC curve is shown in
Figure~\ref{tab:vgoroc}. 

\begin{table}[htb] \centering
  \caption{AD detection results for the VGO data representation
scheme.}
  \label{tab:vgoresults}\vspace*{1ex}
  \begin{tabular}{llll} \hline 
    \multicolumn{2}{c}{AD} & \multicolumn{2}{c}{non-AD}\\ \hline 
    Accuracy$_\mu$ & 0.812 & Accuracy$_\mu$ & 0.714 \\ 
    Precision$_\mu$ & 0.765 & Precision$_\mu$ & 0.769 \\ 
    Recall$_\mu$ & 0.812 & Recall$_\mu$ & 0.714 \\ 
    $F_{1,\mu}$ & 0.788 & $F_{1,\mu}$ & 0.741 \\ 
    Precision$_M$ & 0.667 & Precision$_M$ & 0.792 \\ 
    Recall$_M$ & 0.722 & Recall$_M$ & 0.729 \\ 
    $F_{1,M}$ & 0.685 & $F_{1,M}$ & 0.721 \\ \hline
    \multicolumn{4}{c}{Overall accuracy (LOOCV): 0.811} \\\hline
  \end{tabular}
\end{table}

\begin{figure}[htb] \centering
  \includegraphics[width=\linewidth]{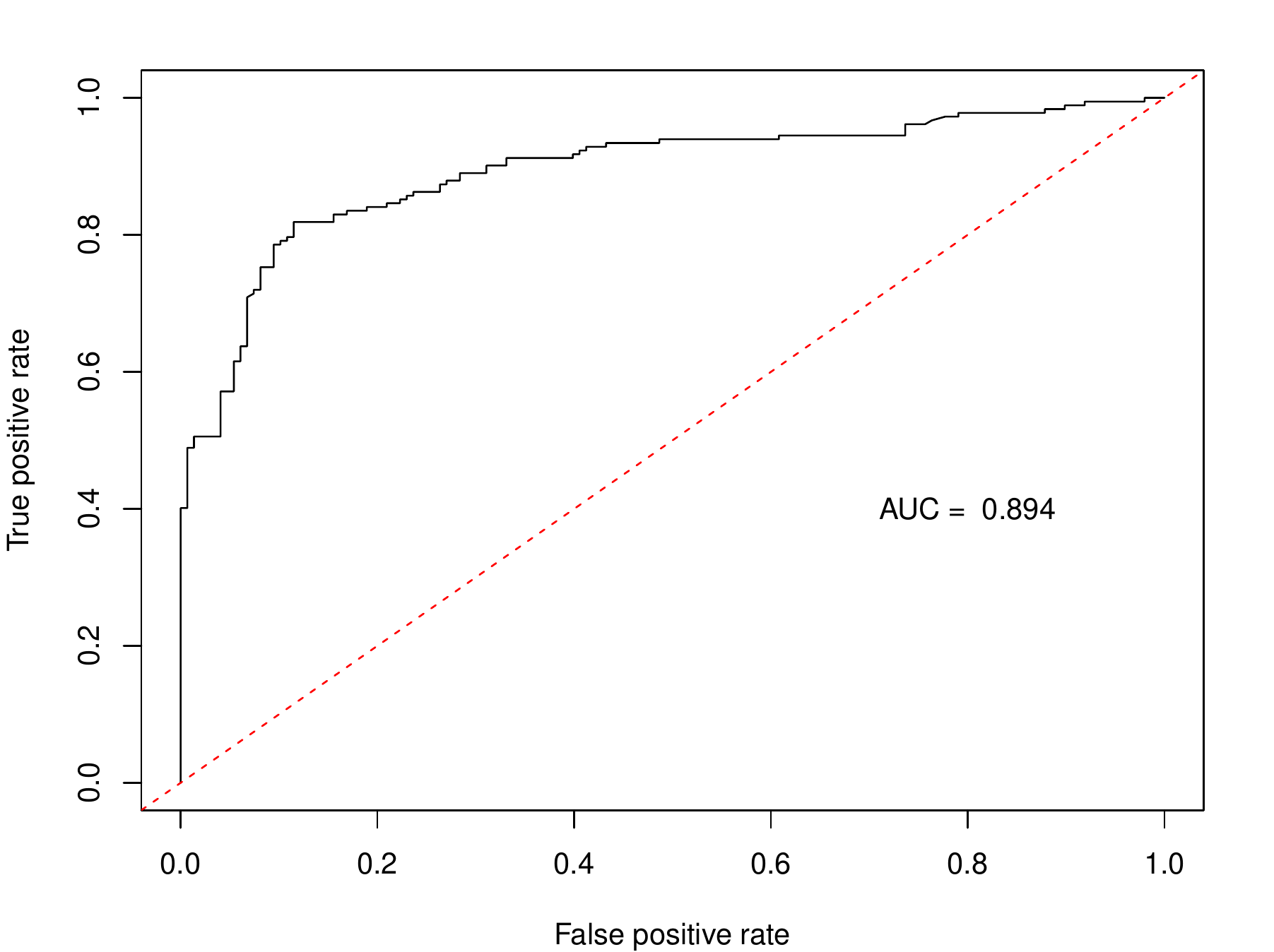}
  \caption{ROC curve for VGS-based classifiers.}
  \label{tab:vgsroc}
\end{figure}

Adding speech rate information (VGS representation) contributed to
further enhancing AD detection, bringing the overall accuracy score to
about 86.5\%. Detailed evaluation metrics are shown in
Table~\ref{tab:vgsresults}.  The ROC curve for the VGS-based
classification approach is shown in Figure~\ref{tab:vgsroc}. It can be
seen that the addition of features for mean and variance of speech
rate ratio over dialogues had the effect of improving classification
trade-offs, particularly reducing the false positives while increasing
the true positives at low threshold cut-offs.

\begin{table}[htb] \centering
  \caption{Results for the VGS data representation scheme.}
  \label{tab:vgsresults}\vspace*{1ex}
  \begin{tabular}{llll} \hline \multicolumn{2}{c}{AD} & \multicolumn{2}{c}{non-AD}\\ \hline 
    Accuracy$_\mu$ & 0.882 & Accuracy$_\mu$ & 0.769 \\ 
    Precision$_\mu$ & 0.833 & Precision$_\mu$ & 0.833 \\ 
    Recall$_\mu$ & 0.882 & Recall$_\mu$ & 0.769 \\ 
    $F_{1,\mu}$ & 0.857 & $F_{1,\mu}$ & 0.800 \\ 
    Precision$_M$ & 0.796 & Precision$_M$ & 0.708 \\ 
    Recall$_M$ & 0.833 & Recall$_M$ & 0.708 \\ 
    $F_{1,M}$ & 0.811 & $F_{1,M}$ & 0.700 \\ \hline
    \multicolumn{4}{c}{Overall accuracy (LOOCV): 0.865} \\\hline
  \end{tabular}
\end{table}

For comparison we ran the same testing procedure using some of the
other classifiers employed in the literature reviewed in
section~\ref{sec:related-work}, namely, logistic regression, na\"ive
Bayes (Gaussian kernel), decision trees (C4.5 algorithm), SVM trained
using sequential minimal optimisation, with a polynomial kernel
\cite{bib:Platt98ftsv}, and random forests \cite{bib:Breiman01rf},
Weka implementation \cite{bib:HallFrankEtAl09w}. The overall (LOOCV)
accuracy figures are shown in Table~\ref{tab:comparison}. There is
little difference in performance between our chosen method (Real
Adaboost) and other methods used in the literature, except for
logistic regression, which underperforms the machine learning
methods. Real Adaboost slightly outperforms SVM and random forests
classifiers, and matches C4.5 decision trees, with a slight advantage
over the latter on the target AD class ($F_m=0.878$ vs.  $F_m=0.872$).

\begin{table}[htb]
  \caption{Compared accuracy results obtained with different
    classification algorithms, on VGS-based datasets.}
  \label{tab:comparison}\vspace*{1ex}
  \centering
  \begin{tabular}{lr}
    \hline
    Classification method & Accuracy (LOOCV) \\
    \hline
    Logistic regression & 75.7\% \\
    Real Adaboost       & 86.5\% \\
    Decision trees      & 86.5\% \\
    SVM                 & 83.7\% \\
    Random forests       & 81.1\% \\
    \hline
  \end{tabular}
\end{table}

Although there is considerable room for improvement upon this level of
classification performance, the levels obtained with these simple
models are comparable to the accuracy of approaches that employ more
detailed linguistic information, which are presumably harder to
acquire in everyday conversational situations, as they would involve a
level of speech recognition accuracy which is beyond the capabilities
of current systems for spontaneous speech in noisy environments.

\section{Conclusion and Further Work}

Dementia prevention and life quality in elderly care are important
societal challenges. Automatic detection of signs of AD in speech can
provide useful tools for the design of technologies for care-giving
and cognitive health monitoring to help address these challenges.

This paper presented initial results of a new method to automatically
recognise the first signs of disrupted communication using dialogue
features. This method obtained an overall accuracy of 0.83, with a
micro F-measure of 0.83 and a macro F-measure of 0.76 on the
classification of patients as ``AD'' and ``non-AD''. Although it is difficult to
compare these results directly to related works
\cite{bib:FraserMeltzerRudzicz16lfidal,guinn2012language}, our accuracy figures are situated
within a similar range, 0.70-0.80, with a smaller discrepancy between
the classification of the two groups, while relying on features that
can be more robustly extracted from spontaneous speech.

Thanks to the increasingly important role of social technology,
longitudinal studies may become richer in terms of the amount of
variables measured, frequency of measurements and places where
measures are taken (living settings), allowing for larger datasets. As
more data are gathered in natural settings, we expect to obtain more
reliable and generalisable results.

There are several linguistic parameters that are promising for the
assessment of cognitive functioning. In current approaches, these features have been typically
extracted from data collected through structured interviews, storytelling or picture
descriptions. The work presented here contributes a new perspective 
to feature extraction by focusing on spontaneous
dialogues. Dialogue processing provides a convenient framework for
the analysis of natural conversations, in which readily available
predictors, such as turn taking behaviour, have already yielded satisfactory
results. We plan to further analyse verbal and non-verbal parameters
to obtain a better characterisations of AD in order to infer
neurosychological assessment results through speech and language
processing, and subsequently to combine such features with actual
neuropsychological evaluations and other relevant variables, building
accurate models to achieve detection of AD at the time of 
onset.

The data set used in the present study has some limitations. Due to its
constraints, the study was performed on a restricted subset of 21+17
sessions. In addition, the interview setting includes a degree of
bias, as the interviewer's objective is to get the patient to perform
a certain task (e.g. description of a picture, driving the discussion)
therefore influencing the patient's speech. In order to mitigate these
limitations, we plan to collect further data in more spontaneous
dialogue in the near future.


\section{Acknowledgements} This research has received funding from the
European Union's Horizon 2020 research and innovation programme under
grant agreement No 769661, SAAM project.  Sofia De la Fuente and
Pierre Albert are supported by the Medical Research Council (MRC). The
authors would also like to acknowledge Charlene Pope and Boyd H. Davis,
from the Medical University of South Carolina, who host the Carolinas
Conversation Collection, for providing access to the dataset and help
in completing the required procedures.


\section{Bibliographical References}
\label{main:ref}

\bibliographystyle{lrec} 
\bibliography{rapid2018,biblio}

\bibliographylanguageresource{rapid2018}

\end{document}